\documentclass[12pt,a4paper]{article}

\usepackage{mathrsfs}
\usepackage{amsmath,amssymb}
\usepackage{hyperref}
\usepackage{verbatim}
\usepackage{stmaryrd}
\usepackage{amsfonts,bbm,xcolor,graphicx}

\usepackage[a4paper,text={450pt,650pt},centering]{geometry}

\newcommand{\fR}{{\mathbb R}}
\newcommand{\fL}{{\mathbb L}}
\newcommand{\fQ}{{\mathbb Q}}
\newcommand{\fC}{{\mathbb C}}
\newcommand{\fGG}{{\mathbb G}}
\newcommand{\fHH}{{\mathbb H}}
\newcommand{\gen}[1]{\mathfrak{#1}}
\newcommand{\genY}[1]{\widehat{\mathfrak{#1}}}
\newcommand{\alg}[1]{\mathfrak{#1}}
\newcommand{\sls}{\alg{sl}}
\newcommand{\psl}{\alg{psl}}
\newcommand{\gl}{\alg{gl}}

\numberwithin{equation}{section}

\let\ifarxiv=\iftrue     % ARXIV VERSION
\ifx\href\asklfhas\newcommand{\href}[2]{#2}\fi
\newcommand{\lrbrk}[1]{\left(#1\right)}

\newcommand{\ep}{e^{\frac{i}{2}p}}
\newcommand{\en}{e^{-\frac{i}{2}p}}

\newcommand{\epp}{e^{i p}}
\newcommand{\enn}{e^{-i p}}

\newcommand{\el}{\nonumber\\}
\newcommand{\bb}[1]{\mathbb{#1}}

\begin{document}

%%%%%%%%%%%%%%%%%%%%%%%%%%%%%%%%%%%%%%%%%%%%%%%%%%%%%%%%%%%

\begin{titlepage}

\noindent {\hfill{DMUS MP 12/04}}

\noindent {\hfill{NORDITA-2012-26}}

\begin{centering}

\vspace{0.3in}

{\Large {\bf Secret Symmetries in AdS/CFT}} 

\vspace{.2in}

{\bf Proceedings to the Nordita program\\ `Exact Results in Gauge-String Dualities'}

\vspace{.2in}

\includegraphics[width=30mm]{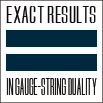}

\vspace{.2in}

{\large Marius de Leeuw${}^{a,1}$, Takuya Matsumoto${}^{b,c,2}$, Sanefumi Moriyama${}^{c,d,3}$, \\Vidas Regelskis${}^{e,f,4}$ and Alessandro Torrielli${}^{g,5}$}

\vspace{.2in}
${}^{a}${\emph{ETH Z\"urich, Institut f\"ur Theoretische Physik, \\Wolfgang-Pauli-Str.\ 27, CH-8093 Zurich, Switzerland}} \\
${}^{b}${\emph{School of Mathematics and Statistics, University of Sydney, \\ NSW 2006, Australia}}\\
${}^{c}${\emph{Graduate School of Mathematics, Nagoya University,\\Nagoya 464-8602, Japan}}\\
${}^{d}${\emph{Kobayashi Maskawa Institute, Nagoya University\\Nagoya 464-8602, Japan}}\\
${}^{e}${\emph{Department of Mathematics, University of York,\\Heslington, York YO10 5DD, UK}}\\
${}^{f}${\emph{Institute of Theoretical Physics and Astronomy of Vilnius University,\\Go\v{s}tauto 12, Vilnius 01108, Lithuania}}\\
${}^{g}${\emph{Department of Mathematics, University of Surrey\\Guildford, GU2 7XH, UK}}\\

%\vspace{.2in}
\footnotetext[1]{{\tt deleeuwm@phys.ethz.ch,}\quad ${}^{2}${\tt tmatsumoto@usyd.edu.au}\quad ${}^{3}${\tt moriyama@math.nagoya-u.ac.jp}\quad ${}^{4}${\tt vr509@york.ac.uk}\quad ${}^{5}${\tt a.torrielli@surrey.ac.uk}}

\vspace{.3in}

{\bf Abstract}

\vspace{.1in}

\end{centering}

We discuss special quantum group ({\it secret}) symmetries of the integrable system associated to the AdS/CFT correspondence. These symmetries have by now been observed in a variety of forms, including the spectral problem, the boundary scattering problem, $n$-point amplitudes, the pure-spinor formulation and quantum affine deformations. 

\end{titlepage}

\tableofcontents

%%%%%%%%%%%%%%%%%%%%%%%%%%%%%%%%%%%%%%%%%%%%%%%%%%%%%%%%%%%%%%%%%%%%%
% Section 1.
%%%%%%%%%%%%%%%%%%%%%%%%%%%%%%%%%%%%%%%%%%%%%%%%%%%%%%%%%%%%%%%%%%%%%

\section{Introduction}
\label{Introduction}

In recent years there has been a remarkable progress towards a proof of the AdS/CFT conjecture \cite{Maldacena:1997re}. The problem of calculating anomalous dimensions in ${\cal{N}}=4$ SYM can be translated into the spectral problem associated to a certain integrable Hamiltonian \cite{Minahan:2002ve}. There are by now extensive reviews on the subject, see for instance \cite{Beisert:2010jr}. From the algebraic viewpoint, it has been possible to rephrase the problem in the language of Hopf algebras and their representation theory.  

The Hopf superalgebra relevant to AdS/CFT is quite unconventional, and, as of today, its properties are only partially understood. It is infinite dimensional, with a structure similar to Yangians \cite{Drin,Chari,Khoroshkin:1994uk,MacKay:2004tc,Molev}. It admits a {\it level zero} given by the centrally-extended $\alg{psl}(2|2)$ Lie superalgebra, and {\it level one} generators giving rise to an infinite dimensional tower\footnote{For Yangians based on Lie superalgebras, see {\it e.g.} \cite{Yao,stuko,Gow,Spill:2008yr}.}. Nevertheless, the actual algebra sits rather outside the standard theory of Yangians, in that it displays an additional symmetry at level one, which is absent at level zero. Were this symmetry present at level zero, it would extend the Yangian to that of $\alg{gl}(2|2)$. However, this is not compatible with the central extension. Moreover, if one starts commuting the new generator with the old ones, one obtains a growth in the algebra which is not completely clear how to control.

At the time it was discovered \cite{Matsumoto:2007rh}, it was unclear whether the secret symmetry was an accidental feature of the choice of vacuum for the spin-chain, or the choice of gauge for the string sigma model. This is because the centrally-extended $\alg{psl}(2|2)$ algebra is intimately linked to those specific choices. More recently, however, there have been observations of the very same mechanism in several ({\it a priori} unrelated) sectors of AdS/CFT. On one hand, this is reassuring that we are not dealing with an accidental problem. On the other hand, even in the light of these new observations, the fundamental nature of the secret symmetry remains unclear, and it is still not known how to consistently embed it into a satisfactory mathematical framework. After all, we might simply have in front of us a new type of quantum group\footnote{P. Etingof, private communication.}.      

In this proceedings\footnote{Based on the talk presented by A.T., Nordita, 15 February 2012.}, we will try and give a survey of the places where the secret symmetry manifests itself. We will start with the spectral problem, where it was originally observed. We will then move to the boundary problem \cite{Regelskis:2011fa}, $n$-point amplitudes \cite{Beisert:2011pn}, the pure-spinor formulation \cite{Berkovits:2011kn} and, finally, to quantum affine deformations \cite{deLeeuw:2011fr}. We will not try and be exhaustive, also due to the fact that some of the above examples were found in the work of others, which we will humbly attempt at reproducing in its salient features.

%%%%%%%%%%%%%%%%%%%%%%%%%%%%%%%%%%%%%%%%%%%%%%%%%%%%%%%%%%%%%%%%%%%%%
% Section 2.
%%%%%%%%%%%%%%%%%%%%%%%%%%%%%%%%%%%%%%%%%%%%%%%%%%%%%%%%%%%%%%%%%%%%%

\section{The algebra}

%%%%%%%%%%%%%%%%%%%%%%%%%%%%%%%%%%%%%%%%%%%%%%%%%%%%%%%%%%%%%%%%%%%%%
% Subsection 2.1
%%%%%%%%%%%%%%%%%%%%%%%%%%%%%%%%%%%%%%%%%%%%%%%%%%%%%%%%%%%%%%%%%%%%%

\subsection{The algebra: level zero}
\label{The Hopf superalgebra: Level zero}

We will start by discussing {the Hopf algebra} based on the Lie superalgebra $A(1,1)=\alg{psl}(2|2)$ with three-fold central extension. We will call this algebra $\alg{psl}(2|2)_c$. Such a large central extension is unique among the basic classical simple Lie superalgebras \cite{IoharaKoga}. The even part of $\alg{psl}(2|2)_c$ consists of $\alg{sl}(2)\oplus \alg{sl}(2)$ and of the space generated by the central elements, which we will denote\footnote{In unitary representations, the central elements $\bb{C}$ and $\bb{C}^{\dag}$ are hermitean conjugate to each other, and so are the supercharges $\bb{Q}$ and $\bb{G}$.} as $\bb{H}$, $\bb{C}$ and $\bb{C}^{\dag}$. Latin indices refer to the first $\alg{sl}(2)$, generated by $\bb{L}_{a}^{\ b}$ subject to $\sum_{a=1}^2 \bb{L}_{a}^{\ a}=0$, greek indices to the second $\alg{sl}(2)$, generated by $\bb{R}_{\alpha}^{\ \beta}$ subject to $\sum_{\alpha=3}^4 \bb{R}_{\alpha}^{\ \alpha}=0$. The fermionic generators will be denoted by $\bb{Q}_{\alpha}^{\ a}$ and  $\bb{G}_{a}^{\ \alpha}$. 

Besides standard $\alg{sl}(2)\oplus \alg{sl}(2)$ commutation relations, one has \cite{Beisert:2005tm}:
\begin{align}
\begin{array}{ll}
\ [\bb{L}_{a}^{\ b},\bb{J}_{c}] = \delta_{c}^{b}\,\bb{J}_{a}-\frac{1}{2}\delta_{a}^{b}\,\bb{J}_{c}, &   \ [\bb{R}_{\alpha}^{\ \beta},\bb{J}_{\gamma}] = \delta_{\gamma}^{\beta}\,\bb{J}_{\alpha} - \frac{1}{2}\delta_{\alpha}^{\beta}\,\bb{J}_{\gamma}, \\
\ [\bb{L}_{a}^{\ b},\bb{J}^{c}] = -\delta_{a}^{c}\,\bb{J}^{b}+\frac{1}{2}\delta_{a}^{b}\,\bb{J}^{c}, & \ [\bb{R}_{\alpha}^{\ \beta},\bb{J}^{\gamma}] = -\delta^{\gamma}_{\alpha}\,\bb{J}^{\beta}+\frac{1}{2}\delta_{\alpha}^{\beta}\,\bb{J}^{\gamma}, \\
\ \{\bb{Q}_{\alpha}^{\ a},\bb{Q}_{\beta}^{\
b}\} = \epsilon_{\alpha\beta}\epsilon^{ab}\,\bb{C},&\ \{\bb{G}^{\ \alpha}_{a},\bb{G}^{\ \beta}_{b}\}=\epsilon^{\alpha\beta}\epsilon_{ab}\,\bb{C}^{\dag}, \\
\ \{\bb{Q}_{\alpha}^{\;a},\bb{G}^{\;\beta}_{b}\} =
\delta_{b}^{a}\,\bb{R}_{\alpha}^{\;\beta} +
\delta_{\alpha}^{\beta}\,\bb{L}_{b}^{\;a}
+\frac{1}{2}\delta_{b}^{a}\delta_{\alpha}^{\beta}\,\bb{H}.&
\end{array}\nonumber
\end{align}
where $\bb{J}$ denotes any {odd} generator with the appropriate index. The elements $\bb{H}$, $\bb{C}$ and $\bb{C}^{\dag}$ commute with all the generators.

The algebra $\alg{psl}(2|2)_c$ can be obtained as a contraction from the simple Lie superalgebra $D(2,1;\alpha)$ (see for instance \cite{Beisert:2005tm,Heckenberger:2007ry,Matsumoto:2008ww,Matsumoto:2009rf}), and for this reason it is sometimes called $D(2,1;-1)$. The Killing form vanishes identically\footnote{This feature is shared by $D(2,1;\alpha)$, $\alg{psl}(n|n)$ and $\alg{osp}(2n+2|2n)$ \cite{Zarembo:2010sg}, and it is crucial for the cancellation of anomalies in the associated string sigma models.}. The algebra admits an $\alg{sl}(2)$ outer automorphism group \cite{Beisert:2006qh}, which is inherited from $A(1,1)$ \cite{Serganova}. This automorphism rotates for instance the three-vector of central charges $(\bb{H}, \bb{C},\bb{C}^{\dag})$ preserving the ``norm" $\bb{H}^2 - \bb{C} \bb{C}^{\dag}$.  

One can put a non-trivial Hopf algebra structure on $\alg{psl}(2|2)_c$ \cite{Gomez:2006va,Plefka:2006ze}. For any $\bb{J}^A \in \alg{psl}(2|2)_c$  
\begin{align}
\label{coprodot}
\Delta (\bb{J}^A) &= \bb{J}^A \otimes \mathbbmss{1} + e^{i [[A]] p} \otimes \bb{J}^A, \el
\Delta (e^{i p})   &= e^{i p} \otimes e^{i p}, 
\end{align}
where $p$ is central. The additive quantum number $[[A]]$ equals $0$ for generators in $\alg{sl}(2)\oplus \alg{sl}(2)$ and for $\bb{H}$, $\frac{1}{2}$ for $\bb{Q}_{\alpha}^{\ a}$, $-\frac{1}{2}$ for $\bb{G}^{\alpha}_{a}$, $1$ for $\bb{C}$ and $-1$ for $\bb{C}^\dag$.
The above coproduct can be easily shown to be a Lie algebra homomorphism. The corresponding counit and antipode are straightforwardly
derived from the Hopf algebra axioms \cite{Plefka:2006ze}.  

The R-matrix $R$ of \cite{Beisert:2005tm} renders this Hopf algebra quasi-cocommutative (see also \cite{Torrielli:2007mc,Arutyunov:2006yd}). One has $R \in U(\alg{psl}(2|2)_c) \otimes U(\alg{psl}(2|2)_c)$, with $U(\alg{psl}(2|2)_c)$ the universal enveloping algebra of $\alg{psl}(2|2)_c$, such that 
\begin{align} \label{invar}
\Delta^{op}(\bb{J}^A) \, R = R \, \Delta(\bb{J}^A) ~.
\end{align}
Since $\Delta (\bb{C})$ is central in $U(\alg{psl}(2|2)_c) \otimes U(\alg{psl}(2|2)_c)$, one must have
\begin{align} \label{coco}
\Delta^{op} (\bb{C}) \, R \, = \, R \, \Delta (\bb{C}) \, = \, \Delta (\bb{C}) \, R \qquad \implies \qquad
\Delta^{op} (\bb{C}) \, = \, \Delta (\bb{C})
\end{align}
(analogously for $\bb{C}^\dag$).
This is guaranteed by the conservation of total momentum in the scattering \cite{Beisert:2005tm}:
\begin{align} \label{interpr}
e^{i p} = \bb{C} \, \, + \, \mathbbmss{1} \qquad\text{and}\qquad
e^{- i p} = \bb{C}^\dag \, \, + \, \mathbbmss{1}.
\end{align}
With these conditions,
\begin{align}
\Delta(\bb{C}) = \bb{C} \otimes \mathbbmss{1} + \mathbbmss{1} \otimes \bb{C} + \bb{C} \otimes \bb{C} = \Delta^{op} (\bb{C}) ~,
\end{align}
and similarly for $\Delta(\bb{C}^\dag)$. The conditions (\ref{interpr}) imply the equivalence relation $\bb{C} \bb{C}^\dag + \bb{C} + \bb{C}^\dag =0$ which we will always assume. This already represents a departure from the standard theory of Hopf algebras. 

%%%%%%%%%%%%%%%%%%%%%%%%%%%%%%%%%%%%%%%%%%%%%%%%%%%%%%%%%%%%%%%%%%%%%
% Subsection 2.2
%%%%%%%%%%%%%%%%%%%%%%%%%%%%%%%%%%%%%%%%%%%%%%%%%%%%%%%%%%%%%%%%%%%%%

\subsection{The algebra: level one}
\label{The Hopf superalgebra: Level one}

The symmetry algebra of the R-matrix contains another set of generators, partners to those described in the previous section. Together with the level zero, these new charges generate an infinite-dimensional Hopf algebra (which we will call $Y$) similar to a Yangian \cite{Beisert:2007ds}. Its Drinfeld's second realization \cite{Dsecond} is given in terms of Cartan generators $\kappa_{i,m}$ and fermionic ladder generators $\xi^\pm_{i,m}$, $i=1,2,3$, $m=0,1,2,\dots$, subject to the following relations \cite{Spill:2008tp}:

\begin{align}
\label{relazionizero}
&[\kappa_{i,m},\kappa_{j,n}]=0,\quad [\kappa_{i,0},\xi^+_{j,m}]=a_{ij} \,\xi^+_{j,m}, \el
&[\kappa_{i,0},\xi^-_{j,m}]=- a_{ij} \,\xi^-_{j,m},\quad \{\xi^+_{i,m},\xi^-_{j,n}\}=\delta_{i,j}\, \kappa_{j,n+m}, \el
&[\kappa_{i,m+1},\xi^\pm_{j,n}]-[\kappa_{i,m},\xi^\pm_{j,n+1}] = \pm \frac{1}{2} a_{ij} \{\kappa_{i,m},\xi^\pm_{j,n}\}, \el
&\{\xi^\pm_{i,m+1},\xi^\pm_{j,n}\}-\{\xi^\pm_{i,m},\xi^\pm_{j,n+1}\} = \pm \frac{1}{2} a_{ij} [\xi^\pm_{i,m},\xi^\pm_{j,n}],
\end{align}

\begin{align} \label{relazero}
& i\neq j, \quad n_{ij}=1+|a_{ij}|, \quad Sym_{\{k\}} [\xi^\pm_{i,k_1},[\xi^\pm_{i,k_2},\dots \{\xi^\pm_{i,k_{n_{ij}}}, \xi^\pm_{j,l}\}\dots\}\}=0 ~, \el
& \text{except for} \qquad \{\xi^+_{2,n},\xi^+_{3,m}\} = \fC_{n+m}, \qquad  \{\xi^-_{2,n},\xi^-_{3,m}\} = \fC^\dag_{n+m}.
\end{align}

The non zero entries of the (degenerate) symmetric Cartan matrix $a_{ij}$ are $a_{12}=a_{21}=1$ and $a_{13}=a_{31}=-1$. The second subscript $n$ 
in $\xi^\pm_{i,n}, \kappa_{i,n}$
denotes the level, with $n=0$ providing a subalgebra isomorphic to $\alg{psl}(2|2)_c$. Our choice corresponds to the following Chevalley-Serre presentation of $\alg{psl}(2|2)_c$ \cite{Dobrev:2009bb}:
\begin{align}
\xi^+_{1,0} &= \,\fGG^{\;4}_2~, &\xi^-_{1,0} &= \,\fQ^{\;2}_4~, &\kappa_{1,0} & = -\fL_1^{\;1} - \fR^{\;3}_3 + \tfrac{1}{2}\fHH~, \el
\xi^+_{2,0} &= i\fQ^{\;1}_4~, &\xi^-_{2,0} &= i\fGG^{\;4}_1~, &\kappa_{2,0} & = -\fL_1^{\;1} + \fR^{\;3}_3 - \tfrac{1}{2}\fHH~, \el
\xi^+_{3,0} &= i\fQ^{\;2}_3~, &\xi^-_{3,0} &= i\fGG^{\;3}_2~, &\kappa_{3,0} & = \;\fL_1^{\;1} - \fR^{\;3}_3 - \tfrac{1}{2}\fHH~.
\end{align}
The generators $\fC_{n}$ and $\fC^\dag_{n}$ are central in $Y$ for all $n$. 

The level one coproduct compatible with (\ref{coprodot}) is quite cumbersome and can be found in the literature. As for ordinary Yangians, it is enough to specify the coproduct at level $0$ and $1$. Recursive use of the defining relations and of the algebra-homomorphism property fixes the coproduct for all other levels. 

One now understands the R-matrix and all generators as living in the universal enveloping algebra of $Y$, denoted as $U(Y)$. The coproduct for the central elements $\fC_{1}$ and $\fC^\dag_{1}$ is non-trivial, although central in $U(Y) \otimes U(Y)$. Following the same argument as in the previous section, quasi-cocommutativity implies $\Delta (\fC_{1}) = \Delta^{op} (\fC_{1})$, and the same for $\Delta (\fC_{1}^\dag)$, hence extra constraints to be added to (\ref{interpr}). We will from now on always assume (\ref{interpr}) and these new level one constraints (which we call {\it hatted constraints}), departing even further from the standard theory of Yangians. 

%%%%%%%%%%%%%%%%%%%%%%%%%%%%%%%%%%%%%%%%%%%%%%%%%%%%%%%%%%%%%%%%%%%%%
% Subsection 2.3
%%%%%%%%%%%%%%%%%%%%%%%%%%%%%%%%%%%%%%%%%%%%%%%%%%%%%%%%%%%%%%%%%%%%%

\subsection{Representations}
\label{Representations: Level zero}

We can use an $\alg{sl}(2)$ outer automorphism to turn $(\bb{H}, \bb{C},\bb{C}^{\dag})$ into $(\bb{H}',0,0)$, corresponding to the Lie superalgebra $\alg{sl}(2|2)$. In turn, $\alg{sl}(2|2)$ is strictly related to $\gl (2|2)$, whose reps we will now describe\footnote{We will follow \cite{Arutyunov:2009pw} in this section.}. 

The paper \cite{zhang-2005-46} (see also \cite{Palev:1990wm} and \cite{Kamupingene:1989wj}) explicitly constructs all
finite-dimensional irreps of $\gl (2|2)$. Generators of $\gl (2|2)$ are denoted by
$E_{ij}$, satisfying %
\begin{align}
[E_{ij},E_{kl}] = \delta_{jk} E_{il} - (-)^{(d[i]+d[j])(d[k]+d[l])}
\delta_{il} E_{kj} ~. 
\end{align}
Indices $i,j,k,l$ run from $1$ to $4$,
and the fermionic grading is assigned as $d[1]=d[2]=0$,
$d[3]=d[4]=1$. The quadratic Casimir of this algebra is $C_2 =
\sum_{i,j=1}^4 (-)^{d[j]} E_{ij}E_{ji}$. Finite dimensional irreps
are labelled by two half-integers $j_1,j_2 = 0,\frac{1}{2},...$,
and two complex numbers $q$ and $y$. These numbers correspond to the eigenvalues of the Cartan elements on highest weight
states $|\omega\rangle$, defined by
\begin{align} \label{zghw} 
H_1 |\omega\rangle &= (E_{11}-E_{22}) |\omega\rangle = 2 j_1 |\omega\rangle, 
& H_2 |\omega\rangle &= (E_{33}-E_{44}) |\omega\rangle = 2 j_2 |\omega\rangle, \el
I |\omega\rangle &= \sum_{i=1}^4 E_{ii} |\omega\rangle = 2 q |\omega\rangle,
& N |\omega\rangle &= \sum_{i=1}^4 (-)^{[i]} E_{ii} |\omega\rangle = 2 y |\omega\rangle, \nonumber
\end{align}
and
\begin{align}
 E_{ij}|\omega\rangle = 0, \qquad \forall \, i<j ~. 
\end{align}
The generator $N$ (to be identified with the level zero generators $\mathbb{B}$ or $\mathfrak{B}$ of the following sections) never appears on
the right hand side of the commutation relations, and is
defined up to the addition of a central element $\beta I$, with
$\beta$ a constant (which we will drop). We
can consistently mod $N$ out, and obtain $\sls
(2|2)$, the algebra of supertraceless $2|2 \times 2|2$ matrices. Further modding out of the center $I$ produces
the simple Lie superalgebra $\psl(2|2)$. Its representations can
be understood as that of $\alg{sl}(2|2)$ at $q=0$
\cite{Gotz:2005ka}.

Irreps of $\gl (2|2)$ are divided into typical
{\it (long)}, with generic values of $j_1,j_2,q$ and dimension $16(2 j_1 +1)(2 j_2 +1)$, and
atypical {\it (short)}, for which special relations are satisfied by
the labels, namely $\pm q = j_1 -
j_2$ and $\pm q = j_1 + j_2 +1$. When these relations are
satisfied, the dimension of the representation is smaller than $16(2 j_1 +1)(2 j_2 +1)$. 

The $4$-dimensional {\it fundamental} representation
\cite{Beisert:2005tm} corresponds to $j_1=\frac{1}{2},j_2=0$ and $q=\frac{1}{2}$. The {\it symmetric bound-state} reps
\cite{Dorey:2006dq,Chen:2006gp,Chen:2006gq,Roiban:2006gs,Beisert:2006qh,Arutyunov:2008zt}
are given by $j_2=0,q=j_1$, with $j_1 =\frac{1}{2},1,...$. The {\it antisymmetric bound-state} reps are given by $j_1=0,q=1+j_2$,
with $j_2 =0,\frac{1}{2},...$. Symmetric and antisymmetric bound-state reps have
dimension $4M$, with $M= 2 j_1$ for symmetric, $M=2(j_2 + 1)$ for antisymmetric. 

An $\sls(2)$ rotation back of such reps provides an explicit matrix
representation of $\alg{sl}(2|2)_c$:
\begin{align}
\fL^{\;b}_a &= E_{a b}~, \quad \forall \, \, a \neq b~,
& \fR^{\;\beta}_\alpha &= E_{\alpha \beta} ~,\quad
\forall \, \, \alpha \neq \beta~, \el
\fQ^{\;a}_\alpha &= a \, E_{\alpha a} + b \, \epsilon_{\alpha \beta} \epsilon^{a b} E_{b \beta} ~,
& \fGG^{\;\alpha}_a &= c \, \epsilon_{a b} \epsilon^{\alpha \beta}
E_{\beta b} + d \, E_{a \alpha} ~, 
\end{align}
subject to the constraint
\begin{align}
 ad - bc = 1. 
\end{align}
Diagonal generators are automatically
obtained by commuting positive and negative roots. 
 
We still need to impose the constraints (\ref{interpr}). This results in further conditions on $a,b,c,d$ which altogether define a certain algebraic curve (details can be found in the literature).

All short representations can be extended to matrix evaluation representations of $U(Y)$ for which (\ref{invar}) holds in $U(Y) \otimes U(Y)$ \cite{Beisert:2007ds,Spill:2008tp}. In particular, the constraints (\ref{interpr}) and the hatted constraints are satisfied, moreover
\begin{align}
\Delta^{op} (\bb{C}_n) \, = \, \Delta (\bb{C}_n)\,, \qquad \Delta^{op} (\fC^\dag_{n}) \, = \, \Delta (\fC^\dag_{n}) \,,
\end{align}
{for all $n$} in these representations. \textit{Evaluation} here means that the level $n$ generators are obtained by multiplying the corresponding level zero matrices by certain polynomials of degree $n$ in a spectral parameter $u$ \cite{Spill:2008tp}. The hatted constraints fix $u$ to be a function of the eigenvalues of the level zero central charges (basically, a function of the momentum $p$) \cite{Beisert:2007ds}. 

%%%%%%%%%%%%%%%%%%%%%%%%%%%%%%%%%%%%%%%%%%%%%%%%%%%%%%%%%%%%%%%%%%%%%
% Section 3.
%%%%%%%%%%%%%%%%%%%%%%%%%%%%%%%%%%%%%%%%%%%%%%%%%%%%%%%%%%%%%%%%%%%%%

\section{Secret symmetry}

%%%%%%%%%%%%%%%%%%%%%%%%%%%%%%%%%%%%%%%%%%%%%%%%%%%%%%%%%%%%%%%%%%%%%
% Subsection 3.1.
%%%%%%%%%%%%%%%%%%%%%%%%%%%%%%%%%%%%%%%%%%%%%%%%%%%%%%%%%%%%%%%%%%%%%

\subsection{Secret symmetry of the R-matrix} 

After computing the element R for all short representations and verifying that it satisfies the Yang-Baxter equation \cite{Beisert:2005tm,Arutyunov:2008zt,Arutyunov:2009mi}, one notices that it automatically solves the additional equation \cite{Matsumoto:2007rh,Beisert:2007ty,deLeeuw:2010nd}
\begin{align} \label{bb}
\Delta^{op} (\widehat{\bb{B}}) \, R = R \, \Delta (\widehat{\bb{B}}),
\end{align}
with
\begin{align}
\Delta (\widehat{\bb{B}}) &= \widehat{\bb{B}} \otimes \mathbbmss{1} + \mathbbmss{1} \otimes \widehat{\bb{B}}
+ \big( e^{\frac{i}{2}p} \, \bb{G}^{\;\alpha}_{a} \otimes \bb{Q}^{\;a}_{\alpha} + e^{-\frac{i}{2}p} \,\bb{Q}^{\;a}_{\alpha} \otimes \, \bb{G}^{\;\alpha}_{a} \big), \el
& \qquad \quad \widehat{\bb{B}} = B_0 \, \, diag(1,1,...,1,-1,-1...,-1). \label{eqn;SS}
\end{align}
The ``$1$"s run over the bosonic subspace of the representation module, the ``$-1$"s over the fermionic subspace, and $B_0$ is a certain function of $p$. One notices that $\widehat{\bb{B}}$ is not supertraceless, unlike all the generators of $Y$. 

One is tempted to postulate an extension of the symmetry algebra to the Yangian of $\alg{gl}(2|2)$. However, this is not possible. The coproduct which one would normally attribute to the level zero partner of $\bb{B}$, namely
\begin{align}
\Delta (\bb{B}) = \bb{B} \otimes \mathbbmss{1} \, + \, \mathbbmss{1} \otimes \bb{B},
\end{align}
with $\bb{B} \propto diag(1,1,...,1,-1,-1...,-1)$, is not a symmetry of the R-matrix. In fact, the R-matrix has entries corresponding to schematic processes
\begin{align}
\mbox{fermion} \otimes \mbox{fermion} \, \, \longrightarrow \, \, \mbox{boson} \otimes \mbox{boson}
\end{align}
and {\it viceversa}, which violate the total fermionic number $\Delta(\bb{B})$. The structure one gets is therefore of a strange {\it indented} Yangian, having a generator at level one with no level zero partners.

From the algebraic viewpoint, the proper way of thinking about the secret symmetry is as an outer automorphism of 
$\alg{psl}(2|2)_c$. As such, it can never be generated from the commutation relations among the $\alg{psl}(2|2)_c$ Yangian 
charges.
Commuting $\widehat{\bb{B}}$ with the level zero supercharges {and taking linear combinations with the coproducts of \cite{Beisert:2007ds}} produces other supercharges, quite different from those already present at level one. 
For example, the new symmetries generated by \eqref{eqn;SS} are\footnote{The additional index $\pm 1$ in (\ref{secretQG}) and (\ref{nuovo2}) does not denote levels of any sort, rather the type of linear combination one needs in order to achieve these coproducts. It is mostly kept here for historical reasons.}
\begin{align}
\Delta\bb{Q}_{\alpha,+1}^{\; a} & =\bb{Q}_{\alpha,+1}^{\; a}\otimes\mathbbmss{1}+\ep\otimes\bb{Q}_{a,+1}^{\;\alpha}-\tfrac{1}{2}\ep\,\bb{L}_{\alpha}^{\;\gamma}\otimes\bb{Q}_{\gamma}^{\; a}+\tfrac{1}{2}\bb{Q}_{\gamma}^{\; a}\otimes\bb{L}_{\alpha}^{\;\gamma} \el
 & \qquad-\tfrac{1}{2}\ep\,\bb{R}_{c}^{\; a}\otimes\bb{Q}_{\alpha}^{\; c}+\tfrac{1}{2}\bb{Q}_{\alpha}^{\; c}\otimes\bb{R}_{c}^{\; a}-\tfrac{1}{4}\ep\,\bb{H}\otimes\bb{Q}_{\alpha}^{\; a}+\tfrac{1}{4}\bb{Q}_{\alpha}^{\; a}\otimes\bb{H}\,, \el
\Delta\bb{Q}_{\alpha,-1}^{\; a} & =\bb{Q}_{\alpha,-1}^{\; a}\otimes\mathbbmss{1}+\ep\otimes\bb{Q}_{a,-1}^{\;\alpha}-\tfrac{1}{2}\varepsilon_{\alpha\gamma}\,\varepsilon^{ac}\,\epp\,\bb{G}_{c}^{\;\gamma}\otimes\bb{C}+\tfrac{1}{2}\varepsilon_{\alpha\gamma}\,\varepsilon^{ac}\,\en\,\bb{C}\otimes\bb{G}_{c}^{\;\gamma},\el
\Delta\bb{G}_{a,+1}^{\;\alpha} & =\bb{G}_{a,+1}^{\;\alpha}\otimes\mathbbmss{1}+\en\otimes\bb{G}_{a,+1}^{\;\alpha}+\tfrac{1}{2}\en\,\bb{L}_{\gamma}^{\;\alpha}\otimes\bb{G}_{a}^{\;\gamma}-\tfrac{1}{2}\bb{G}_{a}^{\;\gamma}\otimes\bb{L}_{\gamma}^{\;\alpha} \el
 & \qquad+\tfrac{1}{2}\en\,\bb{R}_{a}^{\; c}\otimes\bb{G}_{c}^{\;\alpha}-\tfrac{1}{2}\bb{G}_{c}^{\;\alpha}\otimes\bb{R}_{a}^{\; c}+\tfrac{1}{4}\en\,\bb{H}\otimes\bb{G}_{a}^{\;\alpha}-\tfrac{1}{4}\bb{G}_{a}^{\;\alpha}\otimes\bb{H}\,,\el
\Delta\bb{G}_{a,-1}^{\;\alpha} & =\bb{G}_{a,-1}^{\;\alpha}\otimes\mathbbmss{1}+\en\otimes\bb{G}_{a,-1}^{\;\alpha}+\tfrac{1}{2}\varepsilon_{ac}\,\varepsilon^{\alpha\gamma}\,\enn\,\bb{Q}_{\gamma}^{\; c}\otimes\bb{C}^{\dagger}-\tfrac{1}{2}\varepsilon_{ac}\,\ep\,\varepsilon^{\alpha\gamma}\,\bb{C}^{\dagger}\otimes\bb{Q}_{\gamma}^{\; c} \,.\label{secretQG}
\end{align}
The new supercharges display different spectral parameters in the fermion-boson block {\it vs}.\ the boson-fermion block, and when combined together produce the usual Yangian generators,
\begin{align}
\label{nuovo2}
\widehat{\bb{Q}}^{\;a}_{\alpha} = \bb{Q}_{\alpha,+1}^{\; a} + \bb{Q}_{\alpha,-1}^{\; a} \,,\qquad \widehat{\bb{G}}_{a}^{\;\alpha} = \bb{G}_{a,+1}^{\;\alpha} + \bb{G}_{a,-1}^{\;\alpha} \,.
 \end{align}
The growth of the algebra depends on how many new independent charges are generated by subsequent commutation, and in the present case it is still unclear how to control this growth. However, by allowing non-linearity, one can recast {some of} the commutation relations as follows \cite{deLeeuw:2010nd} (see also \cite{Beisert:2011wq,Beisert:2007ty}): 
\begin{align}
\widehat{\bb{Q}}^{\;a}_\alpha &= +[\widehat{\bb{B}},\bb{Q}^{\;a}_\alpha]
+i(1+e^{-ip})\,\epsilon^{ab}\epsilon_{\alpha\beta}\,\bb{G}^{\;\beta}_b\,,
\el
\widehat{\bb{G}}^{\;\alpha}_a &= -[\widehat{\bb{B}},\bb{G}^{\;\alpha}_a]
-i(1+e^{ip})\,\epsilon^{\alpha\beta}\epsilon_{ab}\,\bb{Q}^{\;b}_\beta~. 
\label{secret_comm}
\end{align}
This means that the secret symmetry works as a level raising operator for the Yangian algebra. 
The relations \eqref{secret_comm} are also compatible with the coalgebra structure.

Shadows of this situation are observed at the level of the classical r-matrix, where the secret generator is needed in order to achieve factorization in terms of a Drinfeld double. In fact, all short representations admit a notion of classical limit in a small parameter $\hbar$ \cite{Torrielli:2007mc,Klose:2006zd,deLeeuw:2008dp,Arutyunov:2009mi}. This limit involves a scaling of the eigenvalues of the central elements, since they depend on $\hbar$. The R-matrix can be Taylor-expanded, the first order $r$ being a solution of the classical Yang-Baxter equation \cite{Etingof,BD1,BD2,BD3,Leites:1984pt,Zhang:1990du,Karaali1,Karaali2}. The element $r$ displays a single pole at the origin in some appropriate classical spectral variables, with residue the quadratic Casimir of $\alg{gl}(2|2) \otimes \alg{gl}(2|2)$. There exists an infinite-dimensional Lie bialgebra \cite{Beisert:2007ty}, formulated purely in abstract terms, which admits R as coboundary structure in these short representations. Its nature is quite unconventional, and its quantization is a fascinating open problem. This Lie bialgebra accomodates a class of generators of the type $\widehat{\bb{B}}$, which appear naturally in the classical limit \cite{Moriyama:2007jt}. The {\it indentation} described above is interpreted in \cite{Beisert:2007ty} as a different redistribution of the classical generators in the two copies of the double. {Similarly, the new supercharges and relations we have described above may be interpreted in the classical framework of \cite{Beisert:2007ty}. The difficulty in quantizing the classical Lie bialgebra, however, still prevents from settling the question of how many of these new charges are genuinely independent and how many are not.

Another observation is that the analysis of \cite{Beisert:2007ty} seems to {signal difficulties in having higher secret symmetries of type $\mathbb{B}$ at {\it even} levels}, by inspection of the classical cobrackets\footnote{We thank N. Beisert and B. Schwab for communication about this point, see also remarks in \cite{Beisert:2011pn}.}. This is somehow echoed in \cite{Berkovits:2011kn}, as we will see later on. Other approaches do not seem to detect such obstacles in principle, which is a sign of how non-trivial a task the complete quantum formulation of this symmetry still remains.}

It is worth noticing that in the opposite (gauge-theory) regime of small `t Hooft coupling, the one-loop R-matrix is a twisted version of the $\alg{gl}(2|2)$ Yangian R-matrix in the fundamental representation. The presence of the secret symmetry at level zero is in this case a one-loop accident.

%%%%%%%%%%%%%%%%%%%%%%%%%%%%%%%%%%%%%%%%%%%%%%%%%%%%%%%%%%%%%%%%%%%%%
% Subsection 3.2
%%%%%%%%%%%%%%%%%%%%%%%%%%%%%%%%%%%%%%%%%%%%%%%%%%%%%%%%%%%%%%%%%%%%%

\subsection{Secret symmetries of the K-matrices}

In a series of works \cite{Hofman:2007xp,Correa:2008av,Ahn:2010xa,MacKay:2010zb,Correa:2011nz,MacKay:2011zs} reflection
$K$-matrices were found for open strings ending on $D3$, $D5$ and $D7$-branes. It was also observed that the secret symmetry manifests itself in some of these reflection matrices \cite{Regelskis:2011fa}.

The $Y=0$ maximal giant graviton is a $D3$-brane wrapping a maximal $S^3\subset S^5$ of the $AdS_5 \times S^5$ background and preserves an $\mathfrak{sl}(2|1)_{L} = \{\bb{L}_{\alpha}^{\;\beta},\;\bb{R}_{1}^{\;1},\;\bb{R}_{2}^{\;2},\;\bb{Q}_{\alpha}^{\;1},\;\bb{G}_{1}^{\;\alpha},\;\bb{H}\}$ subalgebra of the bulk algebra $\alg{psl}(2|2)_c$.
The fundamental reflection matrix, describing the scattering of fundamental
magnons from the boundary, is diagonal and the helicity generator
$\bb{B}$ is a symmetry of it, but there is no secret symmetry at level one. Higher
order reflection matrices are of non-diagonal form and respect
neither $\bb{B}$ nor $\widehat{\bb{B}}$. However, the secret symmetry was found to emerge through additional twisted secret charges, namely 
\begin{align}
\widetilde{\bb{Q}}_{\alpha,+1}^{\;2} & = \bb{Q}_{\alpha,+1}^{\;2}+\tfrac{1}{2}\bb{Q}_{\alpha}^{\;2}\,\bb{R}_{2}^{\;2}-\tfrac{1}{2}\bb{R}_{1}^{\;2}\,\bb{Q}_{\alpha}^{\;1}+\tfrac{1}{2}\bb{Q}_{\gamma}^{\;2}\,\bb{L}_{\alpha}^{\;\gamma}+\tfrac{1}{4}\bb{Q}_{\alpha}^{\;2}\,\bb{H} ~,\el
\widetilde{\bb{Q}}_{\alpha,-1}^{\;2} & = \bb{Q}_{\alpha,-1}^{\;2}-\tfrac{1}{2}\varepsilon_{\alpha\gamma}\,\bb{C}\,\bb{G}_{1}^{\;\gamma} ~,\el
\widetilde{\bb{G}}_{2,+1}^{\;\alpha} & = \bb{G}_{2,+1}^{\;\alpha}-\tfrac{1}{2}\bb{G}_{2}^{\;\alpha}\,\bb{R}_{2}^{\;2}+\tfrac{1}{2}\bb{R}_{2}^{\;1}\,\bb{G}_{1}^{\;\alpha}-\tfrac{1}{2}\bb{G}_{2}^{\;\gamma}\,\bb{L}_{\gamma}^{\;\alpha}-\tfrac{1}{4}\bb{G}_{2}^{\;\alpha}\,\bb{H} ~, \el
\widetilde{\bb{G}}_{2,-1}^{\;\alpha} & = \bb{G}_{2,-1}^{\;\alpha}+\tfrac{1}{2}\varepsilon^{\alpha\gamma}\,\bb{C}^{\dagger}\,\bb{Q}_{\gamma}^{\enskip1} ~,
\end{align}
corresponding to \eqref{secretQG} and constructed using the twisted Yangian algebra \cite{Regelskis:2011fa}.

The mirror model of the $Y=0$ maximal giant graviton \cite{MacKay:2010ey,Palla:2011eu} preserves the
subalgebra $\mathfrak{sl}(2|1)_{R} = \{\bb{R}_{a}^{\;b},\;
\bb{L}_{3}^{\;3},\;\bb{L}_{4}^{\;4},\;\bb{Q}_{3}^{\;a},\;\bb{G}_{a}^{\;3},\;\bb{H}\}$. 
The reflection matrices are diagonal for all bound-state numbers; thus $\bb{B}$ is
a symmetry for all bound-states. This configuration also possesses additional twisted secret charges, but no level one secret symmetry $\widehat{\bb{B}}$ itself. 

The $D5$-brane wraps an $AdS_{4}\subset AdS_{5}$
and a maximal $S^{2}\subset S^{5}$ of the $AdS_5 \times S^5$. The $AdS_4$ part of
the brane defines a $2+1$ dimensional defect hypersurface of the $3+1$
dimensional conformal boundary. This brane preserves a diagonal $\mathfrak{psl}(2|2)_{c}$ subalgebra of the complete bulk algebra $\mathfrak{psl}(2|2)_L\times\mathfrak{psl}(2|2)_R\ltimes\bb{R}^{3}$. There are two inequivalent orientations of the $D5$-brane, horizontal and vertical, that look 
rather different in the scattering theory. 

The reflection from the horizontal $D5$-brane, from the scattering theory point of view, is equivalent to the bulk scattering of two identical magnons with opposite momenta. Therefore the reflection matrix respects the same secret symmetries \eqref{eqn;SS}, \eqref{secretQG} as the R-matrix.

In the case of reflection from the vertical $D5$-brane, the boundary carries a field multiplet transforming in the vector representation of the boundary algebra. The reflection factorizes into a sequence of so-called {\it achiral} boundary reflections and bulk scatterings. This reflection is governed by the {\it achiral} twisted Yangian defined on a three-fold tensor space. Once again, the secret symmetry manifest itself in complete generality, however the corresponding expressions for the secret charges are rather cumbersome and we refer the reader to \cite{Regelskis:2011fa}.

These results show how the twisted Yangians inherit most of the properties of the original Yangians. However, while the reflection from the $D5$-brane is
tightly related to the scattering in the bulk (thus the appearance of
the secret symmetry follows naturally), the role of the secret symmetry
in the reflection from the $Y=0$ maximal giant graviton is not yet
understood.

An even more complicated question is related to the reflection from the
$Z=0$ maximal giant graviton and the $D7$-brane. Such scattering is
governed by the level two twisted Yangian \cite{MacKay:2011av}. One
could expect secret charges to be present in this case as well.
However, addressing this issue would require knowledge of the level two twisted
secret symmetry, which has not been explored so far. This is related to
the open question whether the secret symmetry is present at odd levels only, or instead at
all higher levels starting from level one.

%%%%%%%%%%%%%%%%%%%%%%%%%%%%%%%%%%%%%%%%%%%%%%%%%%%%%%%%%%%%%%%%%%%%%
% Subsection 3.3
%%%%%%%%%%%%%%%%%%%%%%%%%%%%%%%%%%%%%%%%%%%%%%%%%%%%%%%%%%%%%%%%%%%%%

\subsection{Secret symmetry in Amplitudes}

In this section, we summarize the findings of \cite{Beisert:2011pn}, who observed the presence of a mechanism completely analog to the one we have been describing for the spectral problem. This time, it involves tree-level planar $n$-particle color-ordered amplitudes $\mathcal{A}_n$. One describes such amplitudes as functions of spinor-helicity variables $(\lambda_k,\tilde{\lambda}_k,\eta_k)$, $k=1,\ldots,n$, with $\lambda_k,\tilde{\lambda}_k\in\bb{C}^2$ complex conjugate spinors, and $\eta_k\in\bb{C}^{0|4}$ a Grassmann variable encoding flavour. The lightlike momentum of the particle $k$ is given by $p_k=\lambda_k\tilde\lambda_k$.

The $\alg{psu}(2,2|4)$ superconformal symmetry generators $\gen{J}^A$ act on particles
as differential operators $\gen{J}^A_k$, and they annihilate the amplitudes. In \cite{Drummond:2009fd}, an additional set of Yangian generators $\genY{J}^A$ annihilating the amplitudes was found, such that
\begin{align}\label{eq:SCYangDef}
\gen{J}^A = \sum_{i=1}^n\gen{J}_i^A ~,
\qquad
\genY{J}^A = f^{A}_{BC}\sum_{j<k=1}^n\gen{J}_j^B\gen{J}_k^C ~.
\end{align}
$f^{A}_{BC}$ are the $\alg{psu}(2,2|4)$ structure constants.
Because of the cyclicity of color-ordered amplitudes, the symmetry generators have to satisfy specific constraints in order to be well-defined. One has for instance 
\cite{Drummond:2009fd}
\begin{align}\label{eq:cyclfree}
\genY{J}^A_{(2,n+1)} - \genY{J}^A_{(1,n)} 
= f^{A}_{BC}\gen{J}_1^{B}\gen{J}^{C} + f^{A}_{BC}f^{BC}_D \gen{J}_1^D=0 ~.
\end{align}
Indeed, the first term vanishes because $\gen{J}^A$ annihilates the amplitudes,
whereas the second term vanishes because the dual Coxeter number of
$\alg{psu}(2,2|4)$ is zero. Curiously enough, this is the same consistency condition pointed out in \cite{Zarembo:2010sg}.

The discovery of \cite{Beisert:2011pn} is that, in addition to the above mentioned Yangian charges, there is an additional ({\it bonus}) charge annihilating the amplitudes, of the same type as the secret symmetry of the spectral problem. This charge would in principle promote the algebra to $\alg{u}(2,2|4)$, however it is not a symmetry at level zero. In fact, the level zero $\gen{B}= \sum_i \eta^A_{i} \partial/\partial\eta^A_{i}$ would count the total helicity, which is not conserved for MHV amplitudes. 

The bonus symmetry $\genY{B}$ acts as
\begin{align}\label{eq:Bhdef}
\genY{B}
=
\sum_{k=1}^{n-1}\sum_{j=k+1}^n 
\lrbrk{\vphantom{\bar{\gen{S}}^{B,}_j }
\ifarxiv\else\begin{array}{l}\fi
\gen{Q}_k^{\alpha b} \gen{S}_{j,\alpha b} 
-\bar{\gen{Q}}_{k,b}^{\dot{\alpha}}\bar{\gen{S}}_{j,\dot{\alpha}}^b
\ifarxiv\else\eqcr[0.3em]\qquad\fi
-\gen{Q}_j^{\alpha b} \gen{S}_{k,\alpha b} 
+\bar{\gen{Q}}_{j,b}^{\dot{\alpha}}\bar{\gen{S}}_{k,\dot{\alpha}}^b
\ifarxiv\else\end{array}\fi
}.
\end{align}
Also the above charge has to satisfy a cyclicity condition which is the analog of (\ref{eq:cyclfree}). One correspondingly finds two terms \cite{Beisert:2011pn}. One term vanishes due to the amplitudes being superconformal symmetric. The other term is now proportional to the dual Coxeter number of $\alg{u}(2,2|4)$, which is non-zero. However, this surviving contribution is multiplied by the central charge $\gen{C}_1$, which vanishes for all individual particles.

Not too dissimilarly from the case of $\alg{psl}(2|2)_c$, 
the bonus symmetry is an automorphism of $\alg{u}(2,2|4)$ and 
plays the role of a level raising operator:
\begin{align}
[\genY{B}, \alg{Q}^{\alpha b}]=+\widehat{\alg{Q}}^{\alpha b}~, \quad 
[\genY{B}, \alg{S}_{\alpha b}]=-\widehat{\alg{S}}_{\alpha b}~. 
\end{align}
The barred supercharges also satisfy similar relations. 

The authors of \cite{Beisert:2011pn} also prove the invariance under $\genY{B}$ of the Grassmannian integral formula \cite{ArkaniHamed:2009vw} for the leading singularities in tree amplitudes. Moreover, they also show that the secret generator can be consistently corrected to ensure that the conformal anomaly is properly taken into account. In fact, certain distributional contributions arise when the differential operators act on poles of the amplitudes. This violates manifest superconformal invariance, which can be restored by adding suitable length-changing operators. One has for the secret symmetry
\begin{align}\label{eq:ExactSym}
\genY{B}\mathcal{A}_n+\genY{B}^+\mathcal{A}_{n-1}=0~,
\end{align}
with
\begin{align}
\genY{B}^+ = \sum_{k=1}^{n-1}\sum_{j=k+1}^n 
\lrbrk{
\ifarxiv\else\begin{array}{l}\fi
\gen{Q}_k^{\alpha b} \gen{S}^+_{j,\alpha b} 
-\bar{\gen{Q}}_{k,b}^{\dot{\alpha}}\bar{\gen{S}}^{+,b}_{j,\dot\alpha}
\ifarxiv\else\eqcr[0.3em]\qquad\fi
-\gen{Q}_j^{\alpha b} \gen{S}^+_{k-1,\alpha b} 
+\bar{\gen{Q}}_{j,b}^{\dot{\alpha}}\bar{\gen{S}}^{+,b}_{k-1,\dot\alpha}
\ifarxiv\else\end{array}\fi\label{shift}},
\end{align}
$\gen{S}^+$ being certain correction terms to the corresponding supercharges \cite{Beisert:2011pn}.
This corrected charge is then also shown to preserve cyclicity. It is fascinating to notice the emergence of the same effect of length-changing in the symmetry action on amplitudes, as one is familiar with from the spectral problem Hopf algebra \cite{Gomez:2006va,Plefka:2006ze}. On the other hand, the action of the level zero and level one Yangian symmetries on amplitudes can also be seen as an $n$-iterated coproduct along the ``spin-chain"/amplitude.

%%%%%%%%%%%%%%%%%%%%%%%%%%%%%%%%%%%%%%%%%%%%%%%%%%%%%%%%%%%%%%%%%%%%%
% Subsection 3.4
%%%%%%%%%%%%%%%%%%%%%%%%%%%%%%%%%%%%%%%%%%%%%%%%%%%%%%%%%%%%%%%%%%%%%

\subsection{Secret symmetry in the pure spinor formalism}

The $\alg{psl}(2|2)_c$ symmetry of the spin chain appears after choosing
one of the complex scalars as a vacuum, while the original super
Yang-Mills theory has manifest $\alg{psl}(2,2|4)$ symmetry.
It is natural to ask whether the original super Yang-Mills theory
itself also possesses a secret symmetry, embedded in some fashion in the Yangian $Y(\alg{gl}(2,2|4))$.
The answer seems to be affirmative \cite{Berkovits:2011kn}.

Compared with the anomalous dimensions or the $n$-point amplitudes,
which are derived quantities, the statement is more direct in the pure spinor formalism.
It was proposed in \cite{Berkovits:2000fe} that string theory on $AdS_5\times S^5$,
or its dual, the super Yang-Mills theory, can be formulated in
the pure spinor formalism.
In \cite{Berkovits:2011kn}, interestingly, it was shown that one can find a secret
symmetry (only at odd levels) in this formalism.

The fact that the super Yang-Mills theory has a symmetry bigger than the Yangian $Y(\alg{psl}(2,2|4))$ but smaller than
$Y(\alg{gl}(2,2|4))$ may look bewildering.
However, an interesting interpretation was suggested by \cite{Berkovits:2011kn}.
Free Yang-Mills theory, which corresponds to the sigma model at zero AdS radius, preserves the whole $Y(\alg{gl}(2,2|4))$, and is
conjectured to be dual to a topological string.
After we turn on the vertex operator that changes the radius, we break
this secret $\alg{gl}(1)$ symmetry spontaneously, with some of its Yangian
cousins remaining. The resulting {\it indentation} is very reminiscent of the one we have been previously discussing.

In the pure-spinor sigma model, one works with a group variable $g\in PSU(2,2|4)$, and the action is given in terms of 
the right-invariant current
\begin{align} \label{ric}
J = -d g \, g^{-1}~.
\end{align}
Integrability is guaranteed by the existence of a Lax connection $J_\pm (z)$ such that \cite{Mikhailov:2007mr}
\begin{align}
\left[ \partial_+ + J_+(z)\;,\; \partial_- + J_-(z) \right] = 0~.
\end{align}
Non-local conserved charges are generated by the transfer matrix
\begin{align}\label{DefinitionOfT}
T(z) = g(+\infty)^{-1}\; \left(P\;\exp \int_C \left(-J_+(z) d\tau^+ - J_-(z) d\tau^-\right)\right) \; g(-\infty)
\end{align}
upon suitable expansion in the spectral parameter, for some contour $C$.
The observation of \cite{Berkovits:2011kn} is that, although $T(z)$ takes values in $PSU(2,2|4)$, one can lift it to $SU(2,2|4)$ by lifting the group element $g$. At this point, one singles out the central component of the transfer matrix by tracing with the hypercharge (which is indeed conjugated to the identity w.r.t.\ the Killing form of $\alg{u}(2,2|4)$). Namely, one takes $\mbox{Str}(s \, \log T(z)) $ with
\begin{align}
s =\; & \left(
   \begin{array}{cc}
   {\mathbbmss{1}}_{4\times 4} & 0 \cr
   0 & -{\mathbbmss{1}}_{4\times 4}
\end{array} \right)~. \nonumber
\end{align}
This procedure generates an infinite family of {\it bonus} charges, in the same spirit as the previous section, but only at odd levels. The first one has the familiar nonlocal expression
\begin{align} \label{SumOfDoubleAndSingle}
\int\int_{\sigma_1<\sigma_2} [j(\sigma_1)\;,\;j(\sigma_2)] - \int k ~.
\end{align}
The density $j$ defines the level zero global conserved charges of the sigma model.
With an appropriate choice of $k$, the authors of \cite{Berkovits:2011kn} prove the independence of the charge they obtain on the lift and 
on the choice of contour. They also prove its BRST closedeness.
As already remarked, the charge (\ref{SumOfDoubleAndSingle}) takes values in the Lie superalgebra $\alg{su}(2,2|4)$.
To obtain a number, one must trace it with a non-supertraceless element $\xi\in \alg{pu}(2,2|4)$, {\it i.e.}
\begin{align}
I_{\xi} = \mbox{Str}\left( \xi\;\;
(\int\int_{\sigma_1>\sigma_2} [j(\sigma_1),j(\sigma_2)] - \int k ) \; \right)~,
\end{align}
which is why one attributes to it the {\it bonus} character. 

%%%%%%%%%%%%%%%%%%%%%%%%%%%%%%%%%%%%%%%%%%%%%%%%%%%%%%%%%%%%%%%%%%%%
% Subsection 3.5
%%%%%%%%%%%%%%%%%%%%%%%%%%%%%%%%%%%%%%%%%%%%%%%%%%%%%%%%%%%%%%%%%%%%%

\subsection{Quantum deformed secret symmetry}

Closely related to $\alg{psl}(2|2)_c$ is the centrally extended quantum deformed algebra $Q = U_q(\alg{sl}(2|2))_c$ \cite{Beisert:2008tw}. This algebra reduces to $U(\alg{psl}(2|2)_c)$ in the rational limit $q\rightarrow 1$ and the corresponding R-matrix describes a deformed one-dimensional Hubbard chain. Moreover, it admits an affine extension $\widehat{Q}$, from which one can recover $Y$ in the rational limit \cite{Beisert:2011wq}. In this sense, $\widehat{Q}$ can truly be seen as a quantum deformation of the Yangian $Y$. The question then arises whether the secret symmetry has a $q$-deformed analogue. This indeed turns out to be the case \cite{deLeeuw:2011fr}.

Let us first briefly discuss the defining relations of $\widehat{Q}$. It is generated by four sets of Chevalley-Serre generators $K_i\equiv q^{H_i}$, $E_i$, $F_i$ ($i=1,2,3,4$) and two sets of central elements $U_{k}$ and $V_{k}$ ($k=2,\,4$)\footnote{Actually, in explicit representations the central elements appear to be inversely related, \textit{i.e.} $U_4 = U_2^{-1}$ and $V_4 = V_2^{-1}$.} with $U_{k}$ being responsible for the braiding of the coproduct similar to the factors of $e^{ip}$ in \eqref{coprodot}. The symmetric matrix $DA$ and the normalization matrix $D$ associated to the Cartan matrix $A$ are
\begin{align}
&DA={\footnotesize \begin{pmatrix}2 & -1 & 0 & -1\\
-1 & 0 & 1 & 0\\
0 & 1 & -2 & 1\\
-1 & 0 & 1 & 0
\end{pmatrix}},
&& D=\mathrm{diag}(1,-1,-1,-1)~. \label{DA}
\end{align}
The algebra is then defined by the following relations \cite{Beisert:2011wq}:
\begin{align}
 & K_{i}E_{j}=q^{DA_{ij}}E_{j}K_{i}~, &  & K_{i}F_{j}=q^{-DA_{ij}}F_{j}K_{i}~, \el
 & [E_{j},F_{j}\} = D_{jj}\frac{K_{j}-K_{j}^{-1}}{q-q^{-1}}~, &  & [E_{i},F_{j}\}=0,\quad i\neq j,\ i+j\neq6~.
\end{align}
These are supplemented by some additional quadratic and cubic (Serre) relations.
The central elements appear in the quartic Serre relations ($k=2,4$):
\begin{align}\label{eqn;quarticSerre}
 & \{[E_{1},E_{k}],[E_{3},E_{k}]\}-(q-2+q^{-1})E_{k}E_{1}E_{3}E_{k}=g\alpha_{k}(1-V_{k}^{2}U_{k}^{2})~, \el
 & \{[F_{1},F_{k}],[F_{3},F_{k}]\}-(q-2+q^{-1})F_{k}F_{1}F_{3}F_{k}=g\alpha_{k}^{-1}(V_{k}^{-2}-U_{k}^{-2})~.
\end{align}
In particular, the central elements on the RHS of \eqref{eqn;quarticSerre} are the analogues of $\mathbb{C},\mathbb{C}^\dag$ and indeed reduce to them in the rational limit.

Remarkably, it turns out that the R-matrix in short symmetric representations \cite{Beisert:2008tw,deLeeuw:2011jr} admits \textit{two} types of secret symmetries. To formulate the exact form of these symmetries, we need multiple versions of the secret symmetry generator similar to \eqref{eqn;SS}
\begin{align}
& \mathbb{B}_{E,F} = B_{E,F} \,diag(1,\ldots,-1\ldots)~,
\end{align}
where $B_{E,F}$ are again explicit functions of the eigenvalues of the central elements. Then it is found that the R-matrix in short symmetric representations respects the following symmetries:
\begin{align}
\Delta \mathbb{B}_E  = & \mathbb{B}_E\otimes \mathbbmss{1} + \mathbbmss{1} \otimes \mathbb{B}_E + \el
& + (U_2^{-1}\!\otimes\!\mathbbmss{1}) (K^{-1}_{123} E_4 \otimes\! \tilde{E}_{123} + K^{-1}_{23}\tilde{E}_{14} \otimes\! \tilde{E}_{23} + K^{-1}_{12} \tilde{E}_{34} \otimes\! \tilde{E}_{12} + K^{-1}_{2}\tilde{E}_{134} \otimes\! E_{2} )+ \el 
& + K^{-1}_{124}E_3 \otimes \tilde{E}_{124} + K^{-1}_3 \tilde{E}_{124} \otimes E_{3} ~, \label{eqn:DBE} \\
\Delta \mathbb{B}_F  = & \mathbb{B}_F\otimes \mathbbmss{1} + \mathbbmss{1} \otimes \mathbb{B}_F +  \el
& + (U_2\!\otimes\!\mathbbmss{1}) (F_4 \otimes\! K_4\tilde{F}_{123} + \tilde{F}_{14} \otimes\! K_{14}\tilde{F}_{23} + \tilde{F}_{34} \otimes\! K_{34}\tilde{F}_{12} + \tilde{F}_{134} \otimes\! K_{134}F_{2} )+  \el
& + F_3 \otimes K_3\tilde{F}_{124} + \tilde{F}_{124} \otimes K_{124}F_{3}~, \label{eqn:DBF}
\end{align}
where $K_{ab} = K_a K_b$ and the non-simple ladder generators  $\tilde{E}_{ab}$ are defined in terms of $E_a,E_b$ and the so-called {{\it right adjoint}} action. Analogous expressions hold also for $F$ and for ladder generators with three indices. For the exact expressions we refer to \cite{deLeeuw:2011fr}.

The first two lines in \eqref{eqn:DBE} and \eqref{eqn:DBF} are both direct generalizations of \eqref{eqn;SS}, since $\tilde{E}_{ab}$ can be thought of as a $q$-deformed commutator of $E_a$ and $E_b$. However, the last term is not present in the undeformed case and appears to be a special feature of the $q$-deformed algebra. These terms from \eqref{eqn:DBE} and \eqref{eqn:DBF} also seem to indicate that this form of the secret symmetry is perhaps not quite universal, due to the asymmetry between the two sets of $\alg{sl}(2)$ generators (\textit{i.e.}\ between indices 1, 3). The exact origin of this discrepancy is not quite clear at the moment, but most likely stems from the corresponding asymmetry present in the bound-state representations that were considered. 

The two secret symmetries can be loosely interpreted as belonging to level ``1'' and ``-1'', respectively. Finally, it is readily checked that in the rational limit, the quantum secret symmetries give rise to the undeformed secret symmetry \eqref{eqn;SS}. Thus, the secret symmetry of the quantum affine model is fully compatible with the one in the undeformed model.

It is interesting to notice how the deformation we have been describing in this section is connected to the so-called Pohlmeyer reduction of the string sigma-model \cite{Grigoriev:2007bu,Hollowood:2009tw,Roiban:2009vh,Hoare:2009fs}, as motivated in \cite{Hoare:2011fj,Hoare:2011wr}. It would be very interesting to investigate the presence of the secret symmetry in the Pohlmeyer-reduced model in terms of non-local classical charges and their quantum lifts%
\footnote{We thank Ben Hoare for discussions on this point.}.

%%%%%%%%%%%%%%%%%%%%%%%%%%%%%%%%%%%%%%%%%%%%%%%%%%%%%%%%%%%%%%%%%%%%% 

\newpage

\paragraph{Acknowledgements.}

\noindent The authors would like to thank everyone who helped in the quest for secret symmetries.

M. dL., S.M., V.R. and A.T. also thank Nordita for warm hospitality during the stay at the program `Exact results in Gauge-String dualities', and the opportunity of giving talks.
M. dL. thanks the Swiss National Science Foundation for funding under the project number 200021-137616. 
S. M. thanks MEXT of Japan for funding under Grant-in-Aids for Young Scientists (B) [\#21740176]. 
V.R. (and A.T. for part of the project) thank the UK EPSRC for funding under grant EP/H000054/1.

%%%%%%%%%%%%%%%%%%%%%%%%%%%%%%%%%%%%%%%%%%%%%%%%%%%%%%%%%%%%%%%%%%%%%
% Bibliography
%%%%%%%%%%%%%%%%%%%%%%%%%%%%%%%%%%%%%%%%%%%%%%%%%%%%%%%%%%%%%%%%%%%%%

\bibliographystyle{unsrt}
\bibliography{SecRev}

%%%%%%%%%%%%%%%%%%%%%%%%%%%%%%%%%%%%%%%%%%%%%%%%%%%%%%%%%%%%%%%%%%%%%

\end{document}